\documentclass[prb,aps,epsf,twocolumn,showpacs,10pt]{revtex4-1}
\usepackage{graphicx,amsfonts,times,bm,amsmath,verbatim,color,array}

\begin{document}
\title{A calculation for polar Kerr effect in high temperature cuprate superconductors}

\author{Girish Sharma$^1$}
\author{Sumanta Tewari$^1$}
\author{Pallab Goswami$^2$}
\author{V. M. Yakovenko$^{2,3}$}
\author{Sudip Chakravarty$^4$}

\affiliation{$^1$ Department of Physics and Astronomy, Clemson University, Clemson, SC 29634\\ $^2$ Department of Physics, CMTC, University of Maryland, College Park, Maryland 20742\\ $^3$ Department of Physics, Joint Quantum Institute, University of Maryland, College Park, Maryland 20742 \\$^4$ Department of Physics and Astronomy, University of California Los Angeles, Los Angeles, California 90095}


\begin{abstract}
A mechanism is proposed for the tantalizing evidence of polar Kerr effect in a class of high temperature superconductors--the signs of the Kerr angle from two opposite faces of the same sample are identical and  magnetic field training  is non-existent. The mechanism does not  break global  time reversal symmetry, as in an  antiferromagnet,  and results in zero Faraday effect. It is best understood in a phenomenological model of bilayer cuprates, such as $\mathrm{YBa_{2}Cu_{3}O_{6+\delta}}$, in which intra-bilayer tunneling nucleates a chiral $d$-density wave such that the individual layers have opposite chirality. Although specific to the chiral  $d$-density wave, the  mechanism  may be  more general to any quasi-two-dimensional orbital antiferromagnet in which  time reversal symmetry is  broken in each plane, but not when averaged macroscopically.

\end{abstract}

\maketitle

\section{Introduction.}

\begin{figure}
\centering
\includegraphics[scale=0.35]{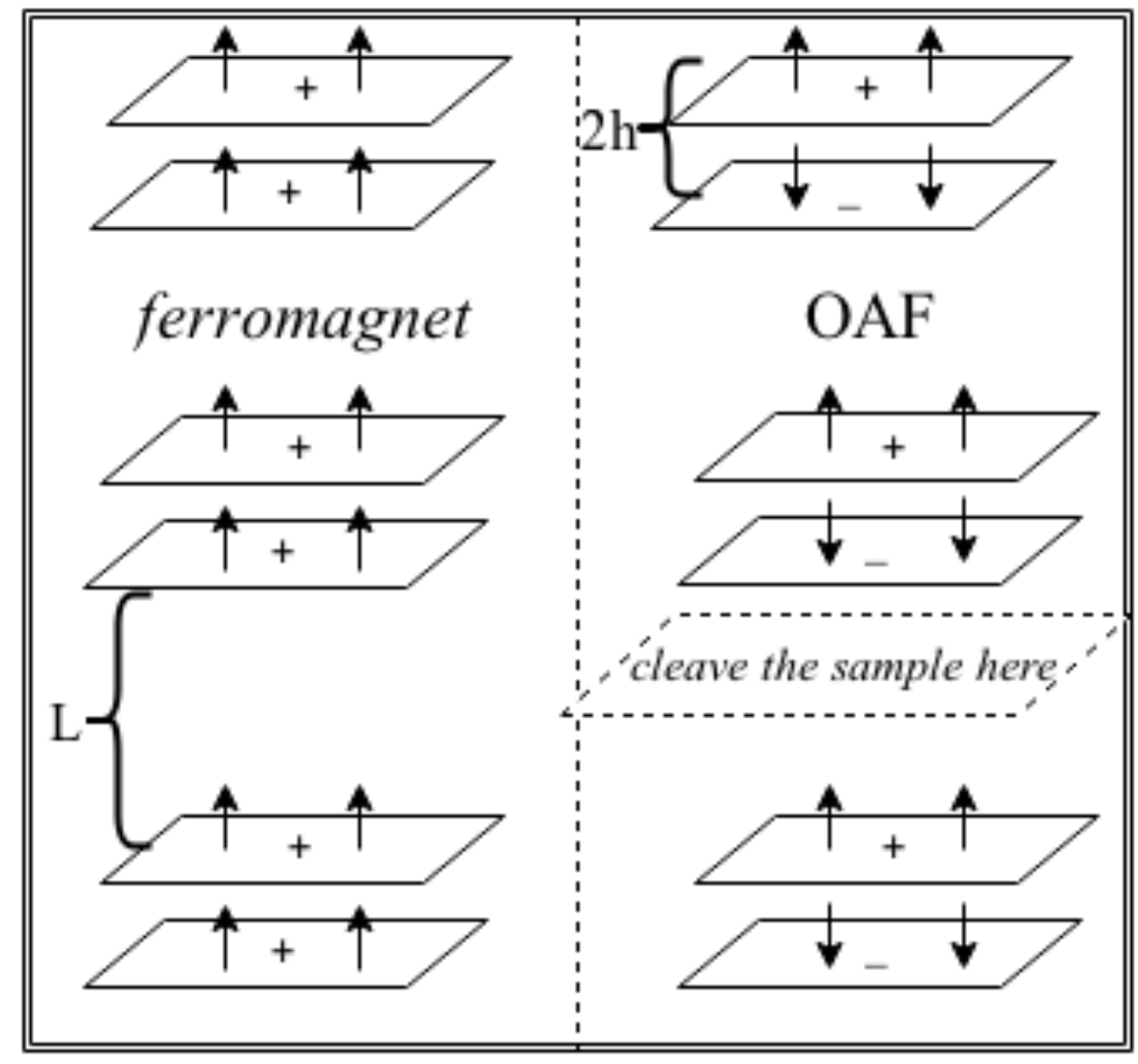}
\caption{\textit{Left panel:} Ferromagnetic (FM) ordering where magnetic moments point in the same direction at each plane, thus resulting in the \textit{opposite} signs of the Kerr angle from the top and bottom surfaces. \textit{Right panel:} Orbital antiferromagnetic (OAF)  ordering, where magnetic moments switch sign on each plane, so the Kerr angle has the \textit{same} sign from both surfaces. In contrast to  FM, magnetic moments in  OAF  point out of the sample at the top and the bottom surfaces, provided the sample is cleaved between the bilayers.}
\label{FM_AFM}
\end{figure}

The origin and nature of the pseudogap phase in the high-$T_c$ cuprate superconductors still remains an unresolved problem ~\cite{Norman:2005,Keimer}. The pseudogap phase, which occurs in the underdoped regime of hole doping, and at temperature range $T^*>T>T_c$, displays many interesting properties including 
various charge, spin, electron nematic, or current ordered states competing with superconductivity~\cite{Norman:2005,Keimer,Varma:1997,Chakravarty:2001,Kivelson:2003,Chakravarty:2013,Wu:2011,Wu:2013,Doiron-Leyraud:2007,
Sebastian:2008,Chang:2012,Ghiringhelli:2012}. Recently a nonzero polar Kerr effect (PKE) has been observed in the pseudogap phase in a number of recent experiments~\cite{Xia:2008,Kapitulnik:2009,He:2011,Karapetyan:2012,Karapetyan:2014}, but with unusual characteristics. The effect measures the angle of rotation of linearly polarized light reflected from a medium at normal incidence and typically signals time reversal symmetry (TRS) breaking in the reflecting medium~\cite{Halperin}. In a  ferromagnetic material, the signs of the polar Kerr angle from two opposite surfaces of the same sample are expected to be different. This is because the net magnetic moment points in the same direction throughout the sample and hence if it points away from the sample on the top surface, it points into the sample on the bottom surface, see Fig.~\ref{FM_AFM}. Moreover, It should be possible to choose (or `train') the direction of the net magnetic moment, and, in turn, the sign of the polar Kerr angle, by cooling the sample in the presence of a magnetic field.

In contrast, in high-${T_{c}}$ superconductors it has been observed that the signs of PKE from the two opposite surfaces of the same sample are identical, and, moreover,  the signal cannot be trained by magnetic field. To account for these puzzling experimental observations, time-reversal invariant models with gyrotropic order were employed recently \cite{Hosur:2013,Orenstein,Mineev:2013,Pershoguba:2013}. However, the concept of gyrotropic order as an explanation for non-zero PKE in the cuprates were subsequently retracted \cite{Mineev:2014,Pershoguba:2014,Hosur:2015} because it does not satisfy Onsager's reciprocity principle in normal reflection that forbids a non-zero PKE in the absence of TRS breaking \cite{Armitage,Fried,Kapitulnik:Physica, Kivelson:2015}.

Here we show that the observations in high-$T_{c}$ can be understood in the framework of a  chiral $d$-density wave state \cite{Tewari:2008} {in the presence of interlayer tunneling}, which is invariant under TRS in the bulk, but can still have a non zero PKE because it is a property of the light reflected from the top surface which breaks TRS locally. Therefore PKE in this mechanism would be insensitive to the changing skin depth of the incoming light at the top surface, while PKE from a bulk order parameter based description would yield a stronger effect for a longer skin depth. The chiral $d$-density wave state is defined by adding  a small $d_{xy}$ component  to the dominant $id_{x^2-y^2}$, i.e. with the combined order parameter $d_{xy}+id_{x^2-y^2}$. The net order parameter  breaks TRS at each CuO plane and results in a non-zero Hall conductivity $\sigma_{xy}$ \cite{Tewari:2008}. The addition of a possible $d_{xy}$ component  could be a result of  microscopic electronic interactions \cite{Nayak:2000}, or   a structural transition that breaks the symmetry between the neighboring plaquettes. In the $id_{x^2-y^2}$ state, by itself, spontaneous currents alternatingly circulate  around plaquettes of the two-dimensional square lattice, thus preserving  the macroscopic TRS, but not  any associated chirality.

We establish that, for our present model, the angle of rotation due to one layer is cancelled by its neighbor, resulting in zero Faraday rotation  of the polarization plane of the transmitted light. However, since
PKE is primarily a surface phenomenon, where the light reflected from the top (or  bottom) surface at normal incidence changes its plane of polarization, there can be a non-zero PKE. Furthermore, since bilayer cuprates usually cleave through the reservoir layers separating the CuO bilayers, the magnetizations at the top and bottom surfaces should point opposite to each other (see Fig.~\ref{FM_AFM}), giving rise to the same sign of the Kerr angle. Finally, since the system as a whole is an OAF, coupling to a small external magnetic field should be small, resulting, most likely, in a small or non-existent magnetic field `training' effect. Importantly, to the best of our knowledge, the scenario presented here is the only one consistent with all the puzzling phenomenology seen in the recent PKE experiments in cuprates. 

It has been argued that much of the phenomenology of the cuprates in the underdoped regime can be unified \cite{Chakravarty:2001,Nayak:2000,Sudip:2001,Sumanta,Sudip:Nature} by making a single assumption that the ordered $id_{x^{2}-y^{2}}$-density wave (DDW) state is responsible for the pseudogap. Moreover, an extensive Hartree-Fock calculation for $id_{x^{2}-y^{2}}$ state has  recently been carried out~\cite{Laughlin:2014}. So far, evidence of magnetism arising from $d$-density wave in neutron or NMR measurements has been controversial. However, the success of the present phenomenological model in explaining PKE must speak in favor of the suggested order parameter.

This paper is divided as follows: in Section II, we introduce the chiral $d$-density wave state order parameter, and calculate the anomalous Hall conductivity of a single layer. In Section III, we discuss the problem of light propagation through a single cuprate layer, and then calculate the Kerr and Faraday responses through the bilayer system in Section IV. We conclude in Section V.

\section{Chiral DDW with interlayer tunneling.}

\begin{figure}
\centering
\includegraphics[scale=0.25]{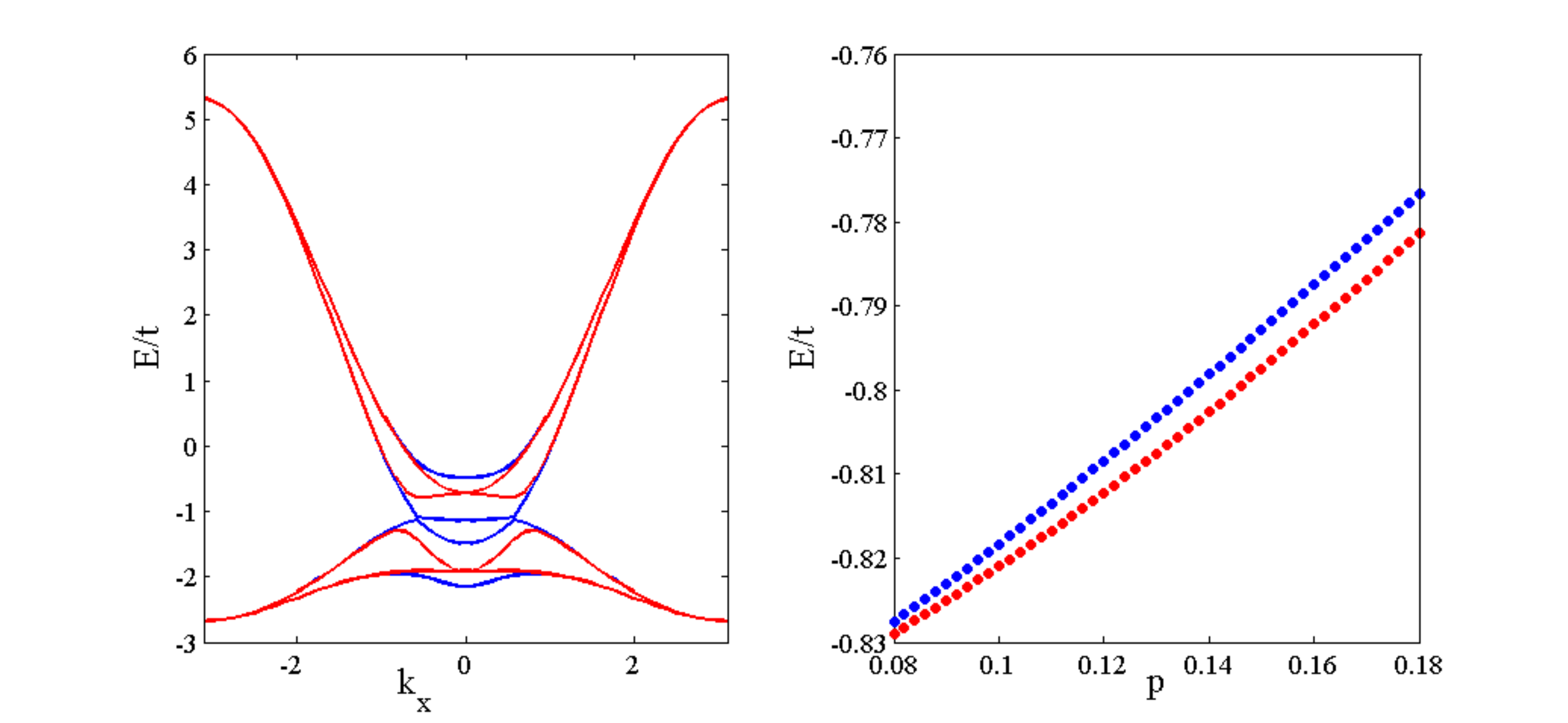}
\caption{(Color online) \textit{Left:} Band dispersion as a function of $\mathbf{k}=(k_x,\pi)$ for the two states in a bilayer: $d+id/d+id$ (blue) and $d+id/d-id$ (red). We utilized $t'=0.33t, W_0=.33t, \Delta_0=0.01W_0/2,$ and $t_{\perp}=0.5t$ all energies measured in the unit of $t$.  \textit{Right:} The ground state energy versus hole doping (from 0.08 to 0.18) indicating that the $d+id/d-id$ state (in red) has lower energy in a bilayer for any given value of hole doping.}
\label{Eigvalues}
\end{figure}

Consider a combination of the density waves, $d_{xy} + id_{x^2-y^2}$, such that the net order parameter is
\begin{equation}
\langle c^{\dagger}_{\mathbf{k}+\mathbf{Q},\alpha} c^{\dagger}_{\mathbf{k},\beta}\rangle \propto [iW_{\mathbf{k}}-\Delta_{\mathbf{k}}]\delta_{\alpha\beta},
\end{equation}
where $c^{\dagger}_{\mathbf{k},\sigma}$ is the electron creation operator of momentum $\mathbf{k}$ and spin $\sigma$, and $\mathbf{Q}=(\pi,\pi)$ is the density wave vector. $W_{\mathbf{k}}$ and $\Delta_{\mathbf{k}}$ correspond to $id_{x^2-y^2}$ and $d_{xy}$  respectively, defined as $W_{\mathbf{k}} = \frac{W_0}{2} (\cos{k_x} - \cos{k_y})$, $\Delta_{\mathbf{k}} = \Delta_0 \sin{k_x}\sin{k_y}$.
We consider a bilayer system where the $id_{x^2-y^2}$ component of the order parameter, i.e.,\ $W_{\mathbf{k}}$ may or may not switch sign between the two layers. The four component mean field Hamiltonian for the system in the basis $\psi^{\dagger}_{\mathbf{k}} =(c^{\dagger 1}_{\mathbf{k}}, c^{\dagger 1}_{\mathbf{k}+\mathbf{Q}}, c^{\dagger 2}_{\mathbf{k}}, c^{\dagger 2}_{\mathbf{k}+\mathbf{Q}})$ takes the following form
\begin{equation}
H(\mathbf{k})=\left( \begin{array}{cccc}
\epsilon_{\mathbf{k}} & g_{1\mathbf{k}} & t_{\perp\mathbf{k}} &  0  \\
g^*_{1\mathbf{k}} & \epsilon_{\mathbf{k}+\mathbf{Q}} & 0 &  t_{\perp\mathbf{k}+\mathbf{Q}} \\
t_{\perp\mathbf{k}} & 0 & \epsilon_{\mathbf{k}} & g_{2\mathbf{k}}    \\
0 & t_{\perp\mathbf{k}+\mathbf{Q}} & g^*_{2\mathbf{k}} & \epsilon_{\mathbf{k}+\mathbf{Q}}   \\
\end{array} \right),
\label{Hk_bilayer}
\end{equation}
where $\epsilon_{\mathbf{k}}$ is the energy dispersion for a two-dimensional square lattice.
\begin{eqnarray}
\epsilon_{\mathbf{k}}=-2t(\cos k_x +\cos k_y) + 4t' \cos k_x\cos k_y,
\label{ek}
\end{eqnarray}
where $t$ and $t'$ are the nearest and next-nearest hopping integrals in the tight-binding Hamiltonian,  $g_{1\mathbf{k}}=iW_{\mathbf{k}}-\Delta_{\mathbf{k}}$ and $t_{\perp\mathbf{k}} = t_{\perp} (\cos {k_x} - \cos {k_y})^2/4$ describes the tunneling between the two layers \cite{Sudip:2001} appropriate for tetragonal systems. The superscript (1,2) on the electron operator in $\psi^{\dagger}_{\mathbf{k}}$ is the layer index. Note that $g_{1\mathbf{k}}=g_{2\mathbf{k}}$ represents a $d+id/d+id$ bilayer configuration and $g^*_{1\mathbf{k}}=g_{2\mathbf{k}}$ is a $d+id/d-id$ configuration. We find that when $g^*_{1\mathbf{k}}=g_{2\mathbf{k}}$ the system is energetically more favorable than the case when $g_{1\mathbf{k}}=g_{2\mathbf{k}}$. This is observed by diagonalizing the Hamiltonian and obtaining the ground state energy for a given doping concentration, as displayed in Fig.~\ref{Eigvalues}. The $d + id$ state spontaneously breaks time-reversal symmetry ($T$) as well as the in-plane reflection symmetry about the principal axes and exhibits anomalous Hall effect with a non-zero value of the Hall conductivity $\sigma_{xy}$. However, the value of $\sigma_{xy}$ reverses sign for the $d - id$ state.  Thus the ground state of the bilayer breaks TRS in each plane, but since the inversion symmetry ($P$) about the mid point between the planes is also broken, the product $PT$ is conserved, allowing the system to have a nonzero polar Kerr effect despite conserving the global TRS and being an OAF ~\cite{Dzyaloshinskii,Canright:1992}. The magnetoelectric effect and PKE in another antiferromagnet Cr$_2$O$_3$ were predicted theoretically in ~\cite{Dzyaloshinskii:1959, Hornreich:1968}, and subsequently observed in experiments~\cite{Astrov:1960, Folen:1961, Rado:1961, Krichevtsov:1993, Krichevtsov:1996}.



The Hall conductance of a single layer described by a $d + id$ mean field Hamiltonian can be calculated using the formalism of linear response theory and Kubo formula \cite{Tewari:2008}. The two-component mean field Hamiltonian describing a $d+id$ density-wave state in the $\psi^{\dagger}_{\mathbf{k}}=(c^{\dagger}_{\mathbf{k}}, c^{\dagger}_{\mathbf{k}+\mathbf{Q}})$ basis is given by:
\begin{eqnarray}
H_s(\mathbf{k})=\left( \begin{array}{cc}
\epsilon_{\mathbf{k}} & g_{\mathbf{k}} \\
g^*_{\mathbf{k}} & \epsilon_{\mathbf{k}+\mathbf{Q}} \\
\end{array} \right).
\label{Hk_single}
\end{eqnarray}
At a finite frequency $\omega$ and in the limit $\mathbf{q}\rightarrow 0$, the anomalous Hall conductivity at any finite temperature is given by \cite{Tewari:2008}:
\begin{equation}
\sigma_{xy}(\omega) = \frac{2e^2}{\hbar}\int\frac{dk^2}{(2\pi)^2}\frac{B(\mathbf{k})f(E_+(\mathbf{k}))-f(E_-(\mathbf{k}))}{w(\mathbf{k})[z-2w(\mathbf{k})][z+2w(\mathbf{k})]},
\label{s_xy}
\end{equation}
where $B(\mathbf{k}) = 4t\Delta_0W_0(\sin^2 k_y +\cos^2 k_y\sin^2 k_x)$ is the Berry curvature, $w(\mathbf{k})$ is the modulus of a three component vector $\mathbf{w}(\mathbf{k})=[-\Delta_{\mathbf{k}}, -W_{\mathbf{k}}, (\epsilon_{\mathbf{k}}-\epsilon_{\mathbf{k}+\mathbf{Q}})/2]$, $f$ is the Fermi distribution function, $\mu$ is the chemical potential, and $z=\omega+i\delta$, with $\delta$ a positive infinitesimal. $E_{\pm}(\mathbf{k})=(\epsilon_{\mathbf{k}}+\epsilon_{\mathbf{k}+\mathbf{Q}})/2\pm w(\mathbf{k})-\mu$ describe the two energy bands obtained by diagonalizing the Hamiltonian in Eq.~(\ref{Hk_single}). The sign of $\sigma_{xy}$ is determined by the sign of the product $\Delta_0W_0$, so the $d\pm id$ states have opposite signs of $\sigma_{xy}$.

\section{Transmission and reflection of light from a single layer.}

We now study propagation of an electromagnetic wave through a layered system with chiral DDW using standard electrodynamics formalism \cite{Dzyaloshinskii,Canright:1992,MacDonald:2010,MacDonald:2011}.  First we consider an electromagnetic wave incident normally on a single two-dimensional layer of a material in the $xy$ plane. The electric field components of the wave in a medium are given by
\begin{eqnarray}
\bar{E} = e^{ikz} \left[ \begin{array}{cc}
E_{+}^{t}  \\
E_-^{t} \\
\end{array} \right] + e^{-ikz} \left[ \begin{array}{cc}
E_+^{r}  \\
E_-^{r} \\
\end{array}\right]
\label{Ea}
\end{eqnarray}
$E_+^{t}$ and $E_-^{t}$ are the transmitted components of right and left circularly polarized (CP) light respectively and similarly $E_+^{r}$ and $E_-^{r}$ are the reflected components, and $k$ is the wavevector. The corresponding magnetic field components can be found using Maxwell's equation, $\bar{k}\times\bar{E} = \omega \bar{H}$. The components of the electromagnetic field satisfy standard electrodynamic boundary conditions at the material layer, which we assume is located at $z=h$: $\bar{E}_{>h} = \bar{E}_{<h}$, $(H_{>h}-H_{<h})_y = -4\pi (\bar{\sigma}\bar{E})_x$, $(H_{>h}-H_{<h})_x = 4\pi (\bar{\sigma}\bar{E})_y$. Note that $\bar{\sigma}$ is a tensor, but $\bar{E}$ is a two component vector. We consider only the Hall components of the surface current and neglect the presence of $\sigma_{xx}$; however, we have confirmed that the presence of a finite $\sigma_{xx}$ due to the chiral DDW Fermi pockets does not change our results qualitatively. We define the scattering matrix which relates the incoming and the outgoing electric field components, $\mathbf{O} = \mathbf{S}\mathbf{I}$,
where the outgoing wave $\mathbf{O}$ and the incoming wave $\mathbf{I}$ are given by,
\begin{equation}
\mathbf{O}=\left[ \begin{array}{cccc}
E_{+,< h}^{r}  \\
E_{-,< h}^{r} \\
E_{+,> h}^{t}  \\
E_{-,> h}^{t} \\
\end{array}\right] \mbox{ and } \mathbf{I}=\left[ \begin{array}{cccc}
E_{+,< h}^{t}  \\
E_{-,< h}^{t} \\
E_{+,> h}^{r}  \\
E_{-,> h}^{r} \\
\end{array}\right],
\end{equation}
and the scattering matrix $\mathbf{S}$ is,
\begin{eqnarray}
\mathbf{S} = \left[ \begin{array}{cc}
\mathcal{R} & \mathcal{T'}  \\
\mathcal{T} & \mathcal{R'}  \\
\end{array}\right]=\left[ \begin{array}{cccc}
R_{++} & R_{+-} & T'_{++} &T'_{+-}  \\
R_{-+} & R_{--} & T'_{-+} &T'_{--}  \\
T_{++} & T_{+-} & R'_{++} &R'_{+-}  \\
T_{-+} & T_{--} & R'_{-+} &R'_{--}  \\
\end{array}\right]
\label{Smatrix}
\end{eqnarray}
This scattering matrix $\mathbf{S}$ describes reflection and transmission of electric field components from the top surface of the slab. We have also defined in Eq.~(\ref{Smatrix}) two-component matrices $\mathcal{R}$, $\mathcal{T'}$, $\mathcal{T}$ and $\mathcal{R'}$, whose components are given by the corresponding block entries. Matching the boundary conditions at $z=h$, we find that
\begin{eqnarray}
R_{++} &=& \frac{e^{ik_>h}\left(1-n^2 -(4\pi\sigma_{xy})^2 +i8\pi\sigma_{xy}\right)}{\left(1 +n \right)^2 + (4\pi\sigma_{xy})^2}  \nonumber \\
R_{--} &=& \frac{e^{ik_>h}\left(1-n^2 -(4\pi\sigma_{xy})^2 -i8\pi\sigma_{xy}\right)}{\left(1+n\right)^2  + (4\pi\sigma_{xy})^2} \nonumber \\
T_{++} &=& \frac{e^{i(k_>-k_<)h}\left(2\left(1+n^2\right) +i8\pi\sigma_{xy}\right)}{\left(1+n\right)^2 + (4\pi\sigma_{xy})^2} \nonumber \\
T_{--} &=& \frac{e^{i(k_>-k_<)h}\left(2\left(1+n^2\right) -i8\pi\sigma_{xy}\right)}{\left(1+n\right)^2 + (4\pi\sigma_{xy})^2} \nonumber,
\end{eqnarray}
where $n$ is the refractive index of the medium. $R'_{++}$, $R'_{--}$, $T'_{++}$,and $T'_{--}$ can be obtained in a similar fashion. The other components  of $\mathbf{S}$ which couple right and left CP components i.e. $R_{+-}, T'_{-+}$ and so on, all vanish. We note that when $\sigma_{xy}\neq 0$, $R_{++}\neq R_{--}$ and $T_{++}\neq T_{--}$ which is a signature of broken time-reversal symmetry.

\section{Polar Kerr and Faraday effects in bilayer chiral DDW.}

\begin{figure}
\centering
\includegraphics[scale=0.35]{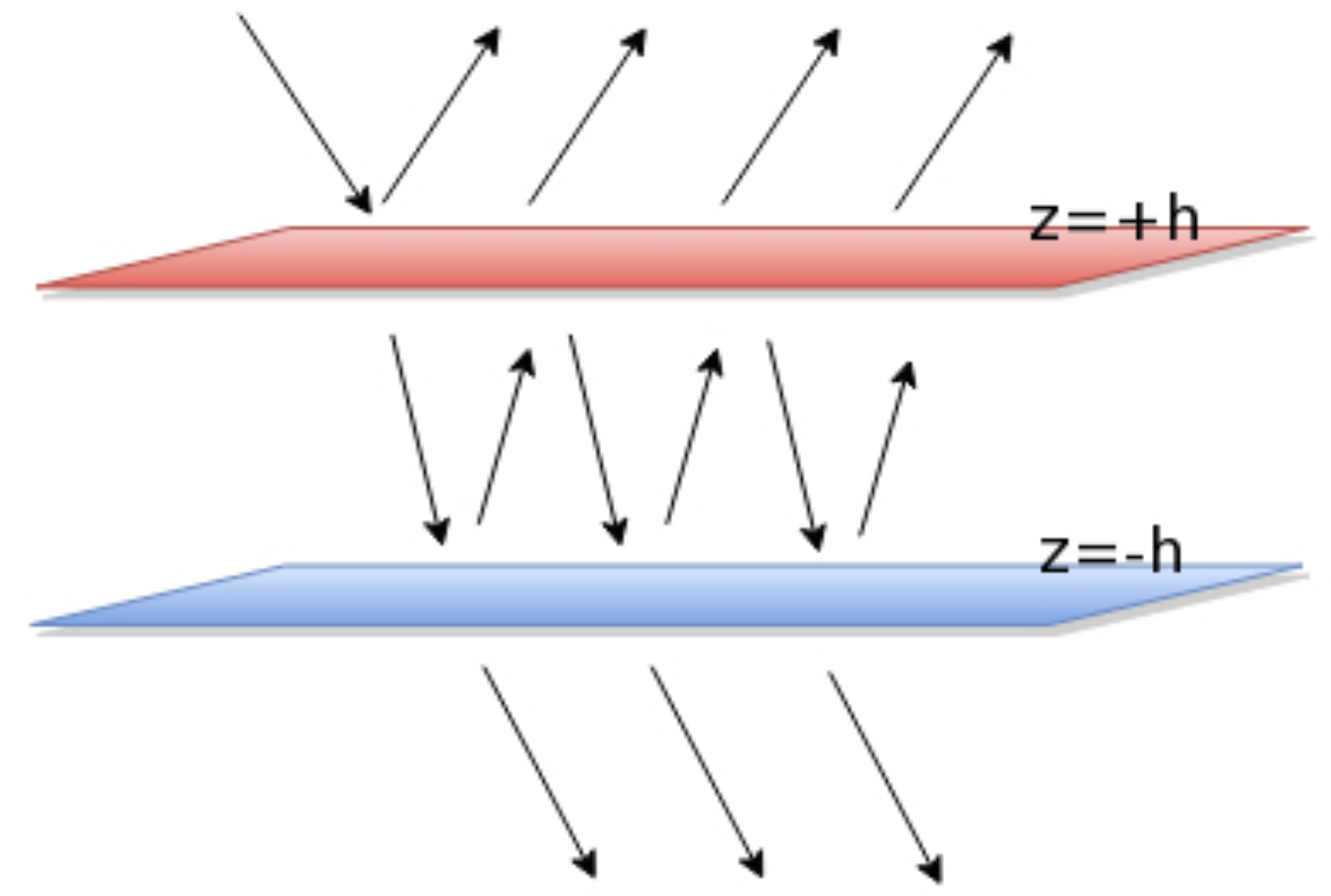}
\caption{(Color online) Schematic diagram showing multiple reflections and transmissions through the top and bottom layers, which we use to calculate the Kerr and Faraday angles. Note that we have assumed normal incidence for the incoming light in our calculations.}
\label{Multiple_refl}
\end{figure}

To discuss scattering from the bilayer, we consider two such interfaces at $z=+h$ and $z=-h$ as depicted in Fig.~\ref{Multiple_refl}. Since even in the presence of the interlayer coupling $t_{\perp}(k)$ the system breaks $P$ and $T$ while $PT$ is conserved, allowing a non-zero PKE \cite{Dzyaloshinskii,Canright:1992}, in the following we will ignore $t_{\perp}(k)$ for simplicity, expecting it to modify our results only quantitatively.  The Hall conductivity $\sigma_{xy}$ reverses it sign at the bottom layer at $z=-h$. One can then appropriately define the scattering matrix elements for the bottom layer taking into account the opposite sign of the Hall conductivity and the position of the bottom plane to be $-h$ instead of $h$. We denote the two-component matrices defined in Eq.~(\ref{Smatrix}) for the top layer by the subscript $T$ and by subscript $B$ for the bottom layer. Thus reflection and transmission through the bilayer as whole are described by tensors $\mathbf{R}$ and $\mathbf{T}$ given by,
\begin{eqnarray}
\mathbf{R} &=& \mathcal{R}_T + \mathcal{T'}_T\mathcal{R}_B (\mathbf{1} - \mathcal{R'}_T\mathcal{R}_B)^{-1} \mathcal{T}_T \nonumber \\
\mathbf{T} &=& \mathcal{T}_B (\mathbf{1} - \mathcal{R'}_T\mathcal{R}_B)^{-1} \mathcal{T}_T
\label{R_T}
\end{eqnarray}
We now switch basis from CP light to linearly polarized (LP) light for convenience of the following discussion. Denoting the electric field of the light incident on the sample by $\bar{E}_I$, $\bar{E}_R=\mathbf{R}\bar{E}_i$ and $\bar{E}_T=\mathbf{T}\bar{E}_i$ give the reflected and the transmitted electric fields. When linearly polarized light is incident on the sample, the Kerr and Faraday angles are determined by the difference between right and left CP light:
\begin{eqnarray}
\theta_F = \frac{1}{2} (\mbox{arg}[E_T^+]-\mbox{arg}[E_T^-])\nonumber \\
\theta_K = \frac{1}{2} (\mbox{arg}[E_R^+]-\mbox{arg}[E_R^-]),
\label{Theta}
\end{eqnarray}
where $E_{R,T}^{\pm} = E_{R,T}^x\pm iE_{R,T}^y$, for $\bar{E}_{R,T} = [E_{R,T}^x, E_{R,T}^y]$. For the bilayer system discussed above, the $\mathbf{R}$ has non-zero off diagonal elements (in LP basis) and $\mathbf{T}$ is diagonal, which is a clear signature of a non-zero Kerr response and the absence of the Faraday effect. (We do not state the analytic expressions for these matrices here, as they are too cumbersome.)  

\begin{figure}
\centering
\includegraphics[scale=0.25]{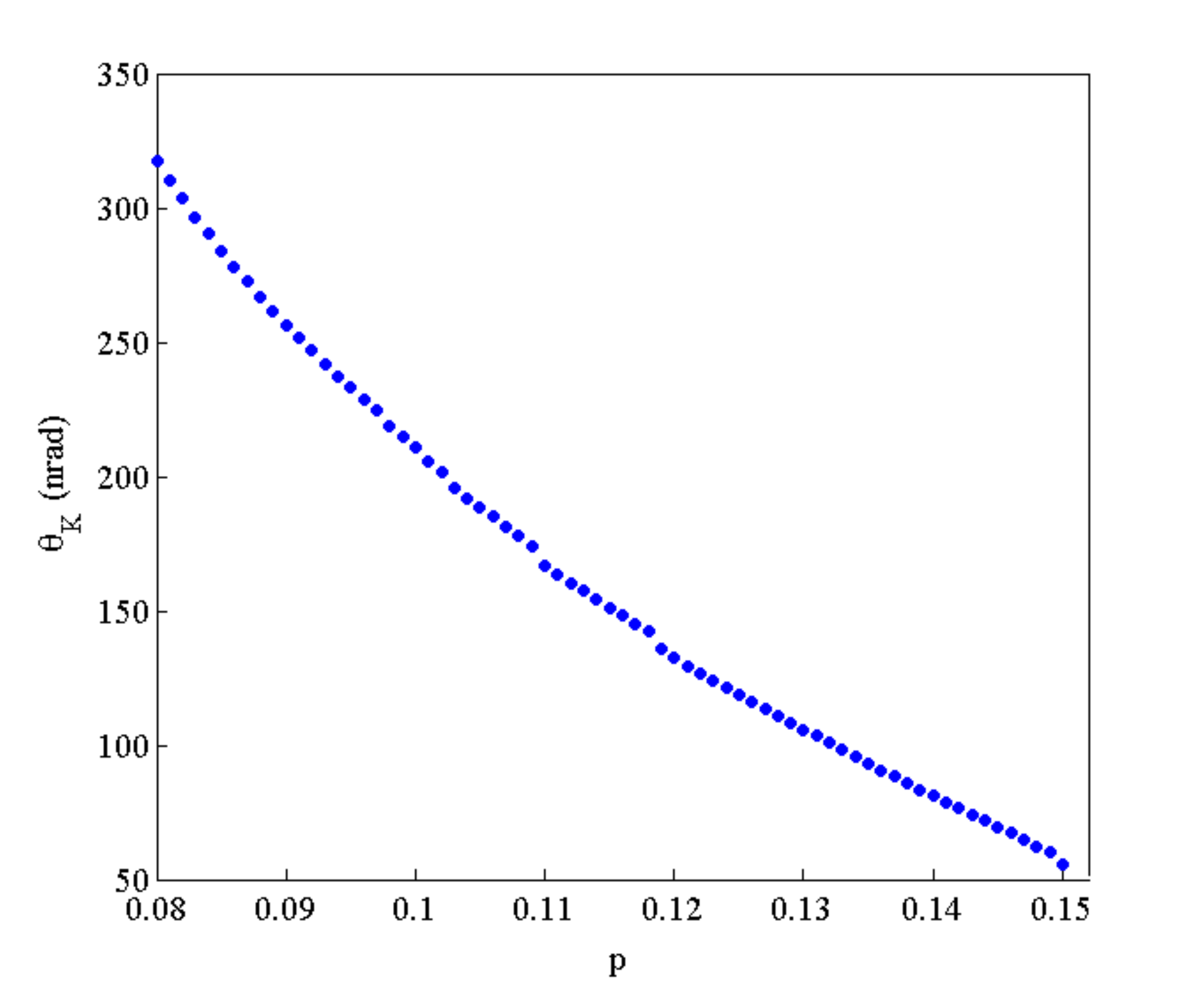}
\caption{Estimated Kerr angle in nrad as function of hole doping $p$. The strength of the $id_{x^2-y^2}$ is assumed to vary with doping as $W(p)=0.1(1-p/p_c)$ eV, where $p_c$ is chosen to be 0.17. The amplitude $\Delta_0$ of the $d_{xy}$ component is assumed to be $1\%$ of $W_0$. While a non-zero PKE is a robust
consequence of our model the precise values of $W_0$, $\Delta_0$, $\theta_K$ in this figure are for illustrative purposes only.}
\label{Kerr_vs_dop}
\end{figure}

We now make a rough estimate for the polar Kerr angle for a bilayer system using Eqs.~(\ref{R_T}) and (\ref{Theta}). Measuring all the energies in units of $t$ we use $t'=0.3$, $\mu=-.9$, $n\approx 1.69$ \cite{Kezuka}, the interlayer distance $2h=3.2$~{\AA}, the strength of the $id_{x^2-y^2}$ component of the order parameter $W_0(p)=0.1(1-p/p_c)$, where $p$ is the hole doping concentration and $p_c=0.17$, $\Delta_0=0.01 W_0/2$ and the frequency of measurement $\omega=1500$ nm.  In Fig.~\ref{Kerr_vs_dop}, we have plotted the polar Kerr angle $\theta_K$ as a function of hole doping and we obtain a non-zero Kerr angle of the order of 100 nrad. The estimated Faraday angle from our formalism turns out to be zero, again from Eqs.~(\ref{R_T}) and (\ref{Theta}).
 Since the chiral DDW with interlayer tunneling is an OAF, the angle of rotation of the plane of polarization of light due to one layer is cancelled by its neighbor, resulting in zero Faraday rotation  of the transmitted light. However, since
PKE is primarily a surface phenomenon, where the light reflected from the top surface changes its plane of polarization, there is a non-zero PKE. Further, since the magnetizations at the top and bottom surfaces should point opposite to each other (see Fig.~\ref{FM_AFM}), the two surfaces give rise to the same sign of the Kerr angle. Finally, since the system as a whole is an OAF, coupling to a small external magnetic field should be small, leading to small or non-existent `training' effect. It is important to note that all of these conclusions are consistent with the phenomenology of the recent PKE measurements in the cuprates.

\section{Conclusions.}

To conclude we considered the chiral DDW state in a bilayer where the sign of the $id_{x^2-y^2}$ component of the order parameter changes between the layers which is an energetically more favorable configuration. This also leads to the reversal of sign of $\sigma_{xy}$ in the bottom layer, thus breaking inversion symmetry. The calculations presented here are consistent with the unusual PKE observed in high-$T_{c}$ materials.
Our calculations, although applied here specifically to the chiral DDW state, are more generally valid for any OAF with TRS broken at each plane.  In Ref.~\cite{Yakovenko:Physica} similar ideas were applied to a tilted loop current model. In addition, the ideas presented here also apply to the bi-axial density wave
recently seen in the pseudogap phase \cite{Chang:2012,Ghiringhelli:2012} if they are accompanied by 
spontaneous currents \cite{Wang:2014,Agterberg:2015,Nandkishore:2014,Gradhand:2015}. However, the theories \cite{Wang:2014,Agterberg:2015,Nandkishore:2014,Gradhand:2015} are currently formulated for a single layer, and it remains to be seen whether they can be generalized to a multilayer model with alternating sign of $\sigma_{xy}$ similar to the present work.

Another important class of high-$T_{c}$ materials  is single layer compounds, such as Bi-2201 ($\mathrm{Bi_{2+x}Sr_{2-x}CuO_{8+\delta}}$) and Hg-1201 ($\mathrm{HgBa_{2}CuO_{4+\delta}}$). Although the detailed results are not yet published, it is known that such materials also show similar PKE, as discussed here~\cite{Kapitulnik}. At the level of order parameter symmetry, there is no difference, in the sense that one can easily envision  CuO-layers alternating between $d+id$ and $d-id$. In addition, recent X-ray measurements indicate that the unit cell in the $c$-direction is doubled, bringing it closer to the bilayer problem. Until  PKE measurements in single layer materials  are published in detail, it is probably prudent to refrain from further speculations.

\textit{Acknowledgement.}
G.S. and S.T. are supported by AFOSR (FA9550-13-1-0045). P.G. is supported by JQI-NSF-PFC. S.C. was supported by a grant from NSF-DMR-1004520.

\end{document}